# Statistical uncertainty in educational experiment on the attenuation of gamma radiation

Mirofora Pilakouta
Department of Physics Chemistry and Material Technology, T.E.I. of Piraeus,
Greece Tel: 2105381583 E-mail:mpilak@teipir.gr

Due to time and financial restrictions in an educational laboratory, we are making compromises, using experimental setups in which limitations and uncertainties are important. In these cases we should pay particular attention to the role of different factors that affect our experiment, in order to achieve the best possible educational outcome and to avoid misconceptions.
In this paper problems related to the use of very low activity source $^{60}$Co in the experiment of measuring the linear attenuation coefficient of gamma rays through matter, will be presented. The role of background radiation in measurements and in the relative statistical uncertainty as well as the role of statistical uncertainty in the choice of representative measurements is discussed. Moreover students' difficulties and misconceptions related mainly to the statistical uncertainty and its connection to measurements overlapping are recorded. An explanation for the possible reasons of these misunderstandings is attempted in order to improve the educational outcome in this experiment.

1.  **Introduction**

The experiment on the attenuation of gamma radiation is usually included in the most introductory physics courses.
A $^{60}$Co radioactive source is sited in front of a small detector (Geiger Muller tube) in an appropriate distance and plates of Pb (or other absorber) are inserted among them. The counting rate as a function of the thickness of the irradiated material is measured and the tasks are: to determine the half-value thickness $X_{1/2}$ (the thickness at which the initial counting rate is reduced by half) the absorption coefficient μ of some materials and to calculate the mass attenuation coefficient from the measured values.

The attenuation of the gamma rays when they pass through an absorber of thickness $x$ is expressed by the exponential law

$$N(x) = N_o\, e^{-\mu x} \qquad (1)$$

(where $N_o$: initial radiation intensity, $x$: the thickness of the material and μ: the linear absorption coefficient). The exponential law is valid under certain conditions i.e monoenergetic point source under narrow beam conditions and absence of background radiation [1].

Furthermore depending on the counting rate yielded by the experimental setup, an appropriate time period for each measurement should be selected in order to have statistically acceptable measurements.

Using experimental setups with low activity source (that commonly happens in an educational lab), the issues concerning background radiation and low statistic are particularly present. In this case the uncertainty of measurements has a very important influence in the design of the experiment. Due to the increased statistical fluctuations, we often get measurements that overlap. This seems to be very difficult for the students to understand. At least, they should have a good understanding in uncertainty of measurements, to understand the reason of the overlapping and the need to get the most representative measurements.

In an introductory physics lab, students are almost never engaged to the design of the experiment. So in practice they have only a theoretical view of the limitations of the experimental setup and the way these limitations affect the measurements uncertainty. Furthermore they rarely care about how uncertainty may vary during the experiment and how this affects the quality of their measurements.

Several authors have studied student's difficulties and misconceptions about the nature and the uncertainty of experimental measurements [3-6]. Their findings show that although the students may successfully calculate mean values and standard deviations or fit data, they have very low understanding of the role of the uncertainty in their measurements and thus they show low ability to evaluate their measurements and form conclusions (4-6).

This paper is focused on the issues of background radiation and low statistics in the experiment on the attenuation of gamma radiation. The role of background radiation in measurements and in the relative statistical uncertainty as well as the role of statistical uncertainty in the choice of representative measurements (under time limitations) is discussed. Moreover (non major in Physics) student's difficulties and misconceptions related mainly to the statistical uncertainty, and its connection to measurements overlapping are recorded. Finally, the first results of a revised instruction sheet that partly engages the students to the design of the experiment are discussed.

## 2. Theoretical approach

2.1 <u>Limitations of the experimental setup</u>

As mentioned above, in low statistic measurements the presence of background radiation is important. The common method to deal with this problem is to obtain firstly a value of the background level and then subtract it from each measurement. Due to the propagation of errors the background subtraction increases the statistical uncertainty of each measurement. Furthermore, as the experimentally recorded net counts $N_{net}$, ($N_{net}$ is proportional to radiation intensity) decreases with the increase of the absorber thickness, the relative statistical uncertainty increases even more and leads frequently in the overlapping of successive measurements. Because of time and experimental setup restrictions we can have only a few numbers of measurements in the above experiment. Thus, a careful selection of the thickness of the absorber for each measurement is needed to get the most representative data. Otherwise, some of the few measurements may have a significant overlap and become meaningless.

To illustrate this situation, let examine the case we often have with our experimental setup in the Physics Laboratory of TEI Piraeus. The "counting rate" from our experimental setup, without any absorbent material, is about 84 counts/min and the background radiation is about 14 counts/min. Thus, we have about 70 counts/min net counting rate.

The time for each measurement is restricted to 4 minutes, so in the available time we can take only 7 measurements in this experiment: two background measurements, a reference measurement (without absorber plates) and four measurements with absorber plates of different thickness.

According to the source net count rate, in a four minute measurement, the initial measurement ($N_{net}$) would be about 280 counts with a relative statistical uncertainty of about 6%. Increasing the thickness of the absorber, the combined statistical uncertainty (σ) may become comparable to the difference ΔN between successive measurements. In order to reduce the possibility of overlapping between two consecutive measurements, the change (ΔN) of the recorded number of counts should be at least twice the statistical uncertainty of the measurements.

Taking into account statistical reasons [1,2], more than 100 net counts are needed in all measurements. Thus the maximum absorber thickness is limited to such a value that the initial radiation intensity is reduced at least to 100 counts.

For this range of absorber thicknesses the uncertainty will vary from about 6 to 12%. Thus to get the most representative measurements we should increase the absorber thickness in our successive measurements nonlinearly in order for the measurements to be approximately equally-spaced in the range 280 and 100 counts and have less possibility to overlap.

2.2 <u>The role of background radiation in the statistical uncertainty of measurements - Measurements overlapping</u>

The contribution of the background radiation in the estimation of combined uncertainty [2] is given in the Appendix.

The relative statistical uncertainty of the net counts of a measurement as a function of the ratio $N_b/N_{o\ net}$ and the reduction factor α of the initial radiation intensity is given by Eq. (4) in the Appendix.

For α=2, the initial intensity of the radiation is reduced by half and the corresponding relative statistical uncertainty becomes:

$$\sigma'_{net/2} = 1.4 \sigma'_{net_0} \sqrt{(1 + \frac{4N_b}{N_{onet}})} \qquad (2)$$

Equation (2) indicates that the relative statistical uncertainty in this case, increases at least 1.4 times in comparison to the relative uncertainty of the measurement without absorber. In addition, the presence of background radiation causes further enlargment of the statistical uncertainty. The importance of background radiation in low statistic measurements is illustrated in Table 1.

Three cases with different number of initial counts **$N_s$ ($N_s$** includes background) are presented. In each case the absolute **σ$_{net}$** and relative **σ'$_{net}$** statistical uncertainty for the initial measurement and $\sigma'_{net/2}$ and $\sigma_{net/2}$ respectively for a measurement with half the initial counts are listed. All the counts correspond to the same period of time, 4 minutes. The background radiation is considered $N_b = 56$ counts.

| $N_s$ | $N_{o\ net}$ | $\sigma'_{neto}$ | $\sigma'_{net}$ | $\sigma_{net}$ | $N_{o\ net}/2$ | $\sigma'_{net/2}$ | $\sigma_{net/2}$ |
|---|---|---|---|---|---|---|---|
| 1000 | 944 | 0.03 | 0,03 | 32 | 472 | 0.05 | 24 |
| 500 | 444 | 0.05 | 0,05 | 24 | 222 | 0.08 | 18 |
| 350 | 294 | 0.06 | 0,07 | 20 | 147 | 0.11 | 16 |

**Table 1:** Absolute and relative statistical uncertainty for three cases with different number of initial counts.

In the first case we have $N_s = 1000$ counts and the range of possible values (with 68% confidence level) is 944 ± 32. This means that a decrease in intensity by 10% (which roughly corresponds to a thickness of 2 mm Pb using source Co) will give $\Delta N \sim 95 > 2\ \sigma_{net}$ and the probability of overlapping of two successive measurements is negligible. However, in the region of $N_{o\ net}/2$, a 10 % intensity change is comparable to twice the $\sigma_{net/2}$.

In the second case with $N_s = 500$ counts, a 10% reduction of the initial net counts is comparable to 2 $\sigma_{net}$. So we should use absorber with a suitable thickness to produce a reduction greater than 10% at the beginning. In the region of $N_{o\ net}/2$, the change between two successive measurements must be at least 16% (Table 1)

The third case corresponds to the case referred in the previous subsection. $N_s = 350$ and in order to have 68% chance to get distinctness between two successive measurements, the initial absorber thickness should reduce roughly 14% the initial counts (~ 2.5 mm Pb) while in the region of $N_{o\ net}/2$, the change must be at least 22% (~ 5 mm Pb). In this case an increasingly larger $\Delta x$ is needed to measure substantial differences between two successive measurements.

In summary, in case there is sufficient count rate or if there is plenty of time to take many measurements, no particular problem will be noticed related to measurements overlapping during successive increase of the thickness x of the absorber. Under low count rate and limited time conditions we should take care to get the most representative measurements. Therefore in this case we should increase the absorber thickness between successive measurements nonlinearly and take into consideration the statistical uncertainty.

### 3. Educational approach

#### 1.1 Student's difficulties and misconceptions associated to measurements uncertainty and measurements overlapping

The measurements overlapping that appears frequently due to the limitations of the experimental setup, has revealed several difficulties and misconceptions related to the uncertainty and its significance in this experiment. Interviewing more than 40 students that faced problems during the acquisition of their data, we have recorded these difficulties and misconceptions but we have also recorded difficulty to link the uncertainty with the overlapping of measurements. Most of our observations are similar to the findings mentioned in educational studies related to students' understanding of measurements [4-6]. In the following we present our observations accompanied with comments and the first results of a revised instruction sheet that partly engages students to the design of the experiment.

- **Origin of the statistical uncertainty when measuring gamma radiation**

   The majority of students do not seem to realize the source of the uncertainty in a counting experiment such as the radiation measurement. They connect the uncertainty ("measurement error") with the counter and not with the random nature of the emitted radiation. They view each reading of N counts as an exact value because it is taken by a digital counter.
   This misconception is possibly related to their little experience with the statistical fluctuation of the nuclear radiation but also because many of them have difficulties to understand the existence of uncertainty in one measurement [5]. Some of them consider that the uncertainty of the counts recorded by the counter is similar to the uncertainty corresponding to the measuring of a length with a ruler (i.e consider as uncertainty the lowest division of the measuring instrument.

- **Uncertainty calculation in gamma radiation experiment**

   More than 80% of the students fail to calculate correctly the statistical uncertainty (absolute and relative) because they underestimate or ignore the background radiation or they do not feel enough confident to use the combined uncertainty. Thus measuring $N_s$ counts they find $N_{net} = N_s - N_b$ but estimate the uncertainty using $\sqrt{N_{net}}$ instead of using $\sigma_{net} = \sqrt{\sigma_{N_s}^2 + \sigma_{N_b}^2}$
   This may be related to the fact that background radiation is not enough underlined in the most introductory textbooks and that in theoretical problems related to the attenuation of radiation, most of the times the background radiation is ignored.

- **The role of uncertainty in measurements interpretation**

   Almost all the students have difficulties to interpret their measurements using the uncertainty associated with them.
   In this experiment, students consider as an expected result the decrease of the recorded counts after any increase of absorber thickness. They also expect to find clearly distinct measurements when increasing the absorber thickness. Moreover they consider that increasing the thickness by ΔX they should always get about the same difference ΔN in recorded counts. As mentioned above, increasing ΔX linearly, it is mostly probable for some of the measurements to overlap significantly (within the range ± σ).
   Students become skeptical if the counter indicates about the same number of counts for two different absorber thicknesses, and they consider it as a conflicting result, if in two successive measurements, they find more counts in the measurement taken with the thicker absorber. For example, in an experiment with initial $N_{net}$=290 counts, using 5 mm of Pb the counter recorded $N_{net}(5)$=205 counts. Adding another 2 mm (thus total thickness 7 mm) the counter recorded $N_{net}(7)$=212 counts. It appeared to be very hard for the students to understand that the above measurement is valid within the statistical uncertainty. When they face the above problem, they consider either that the counter is not working properly or that they have done something wrong and most of the times they repeat the measurement.

Actually students tend to compare measurements as point numbers without taking into account the range (± σ) of possible values within which the true value of the measurement lies. Thus they are not able to see the above example as a case of measurements overlapping.

In the preparatory physics lab, the students do not have the opportunity to clarify aspects like the sensitivity of the experimental setup. In most experiments the range and the step of measurements is determined by the teacher, or a large number of measurements is available so the students can analyze their data without any doubt or consideration about the quality of the measurements.

### 3.2 Extending the objectives of the experiment

In an attempt to improve our students' understanding on the meaning of their measurements in this experiment, we expanded the objectives of the experiment. Apart of using the experimental technique to find the attenuation coefficient of some absorbers, the students are engaged partly to the design of the experiment to realize in practice the importance of measurement uncertainty.

In the above experiment, it's possible to determine the uncertainty of the measurements at the beginning of the experimental process. All that is needed for this is the background and the source counting rate. Thus we inserted in the laboratory worksheet questions with which the students are guided to

- decide for the duration and the number of measurements
- discuss the main sources of uncertainty in their measurements
- discuss how they could reduce the uncertainty in this experiment
- preestimate the range of the uncertainty (using relation (2)) for some representative cases as indicated in Table 1 and use the results for the justification of their measurements or for deciding for the appropriate thickness of the absorber to be used to get the most distinct measurements.

The revised sheet accompanied with personal instructions was tested in 18 students (6 groups of 3 students) during the last semester. The first findings using this revised laboratory instruction sheet shows that at least half of the students show a better understanding of the experimental procedure and of their measurements and admit that the preestimation of the uncertainty helps them to understand better the outcome of the experiment.

## 4. Conclusion and outlook

In this paper issues related to the use of a low activity source in the experimental setup for measuring the linear attenuation coefficient of gamma rays through Pb, were presented. Students' difficulties and misconceptions in this experiment indicates the need for further explanation and clarification of:
- the main sources of uncertainty in measurements and the combined uncertainty
- the significance of the uncertainty to the interpretation of the measurements and the experimental designing

The first findings of a revised instruction sheet that partly engages the students to the design of the experiment seem to be encouraging. The revised instruction sheet in combination to students interviewing will be used in the following semesters for a greater number of students as a step to achieve a better educational outcome from this experiment.

## Appendix

- How the background radiation affects the relative statistical uncertainty of the measurements

Here $N_b$ : are the counts due to background radiation and $\sigma_{N_b} = \sqrt{N_b}$ their statistical uncertainty. The total number of counts is $N_s$ and $\sigma_{N_s} = \sqrt{N_s}$ their statistical uncertainty. The N$_{net}$ is the symbol for the net counts of a measurement $N_{net} = N_s - N_b$ and $\sigma_{net}$ their combined statistical uncertainty

$$\sigma_{net} = \sqrt{\sigma_{N_s}^2 + \sigma_{N_b}^2} = \sqrt{N_s + N_b} = \sqrt{N_{net} + 2N_b} \qquad (1)$$

The relative statistical uncertainty of the net counts $\sigma'_{net}$ is given by:

$$\sigma'_{net} = \frac{\sigma_{net}}{N_{net}} = \frac{1}{N_{net}}\sqrt{N_{net}(1+\frac{2N_b}{N_{net}})} = \frac{\sqrt{N_{net}}}{N_{net}}\sqrt{(1+\frac{2N_b}{N_{net}})} \qquad (2)$$

And setting $\sigma'_{net_0} = \frac{\sqrt{N_{net}}}{N_{net}}$ ( the relative statistical uncertainty in the case of zero background) Eq. (2) becomes

$$\sigma'_{net} = \sigma'_{net_0}\sqrt{(1+\frac{2N_b}{N_{net}})} \qquad (3)$$

Equation (3), shows how the background radiation affects the relative statistical uncertainty of the measurements.

For $N_b \ll N_{net}$ the ratio $\frac{N_b}{N_{net}}$ is very small and approximately $\sigma'_{net} = \sigma'_{net_0}$

When the net recorded counts are low, the background radiation contributes substantially to the overall relative statistical uncertainty

(for example if $\frac{N_b}{N_{net}} = 0.2$ then $\sigma'_{net} = 1.18\sigma'_{net_0}$)

- <u>How is the relative statistical uncertainty changes due to the absorption of the radiation through matter</u>

If $N_{o\ net}$ are the net counts taken with no absorber present and $N_{o\ net}/a$ the net counts taken with an absorber of thickness X$_{1/α}$ (that corresponds to attenuation of the initial intensity by a factor of α ), the corresponding relative statistical uncertainty $\sigma'_{net/a}$ is given by:

$$\sigma'_{net/a} = \frac{\sigma_{net/a}}{N_{onet}/a} = \sqrt{a}\sigma'_{net_0}\sqrt{(1+\frac{2aN_b}{N_{onet}})} \qquad (4)$$

**Acknowledgements**

The author would like to thank all the associates of the physics lab II for helpful discussions and especially Dr. Christos Dedes for his comments and suggestions for this manuscript.

**References**

[1] Harald A. Enge, "Introduction to nuclear physics, p.191,235", 9th Ed.(Addison- Wesley, 1979)
[2] John R. Taylor, "An Introduction to Error Analysis: The Study of Uncertainties in Physical Measurements", 2nd ed. (Univ. Science Books, 1997)
[3] Rebecca Lippmann Kung, "Teaching the concepts of measurement: An example of a concept-based laboratory course", Am.J.Phys. vol.78, no 8, pp.771-777, August 2005.
[4] Marie-Genevieve Sere, Roger Journeaug, and Claudine Larcher. "Learning the statistical analysis of measurement errors", Int. J. Sci. Educ.,vol 15 ,no 4, pp. 427-438, July 1993)
[5] S. Allie, A. Buffler, B. Campbell, F. Lubben, D. Evangelinos, D. Psillos, and O. Valassiades "Teaching measurement in the introductory physics laboratory, " Phys. Teach. Vol.41, pp.394-401, October 2003.
[6] Trevor S. Volkwyn, Saalih Allie, Andy Buffler and Fred Lubben "Impact of a conventional introductory laboratory course on the understanding of measurement" Phys. Rev. S.T – Phys. Edu Res. Vol.4, no.1, pp.010108_1-10 ,May 2008.